\documentclass{JINST}

\title{The performance of the LHCf detector for hadronic showers}

\author{
K.~Kawade$^{1}$, O.~Adriani$^{2,3}$, L.~Bonechi$^{2}$, M.~Bongi$^{2,3}$, G.~Castellini$^{2,4}$, R.~D'Alessandro$^{2,3}$, M.~Del~Prete$^{2,3}$,
M.~Haguenauer$^{5}$, Y.~Itow$^{1,6}$, K.~Kasahara$^{7}$, Y.~Makino$^{1}$, K.~Masuda$^{1}$, E.~Matsubayashi$^{1}$, H.~Menjo$^{8}$, G.~Mitsuka$^{1,3}$, Y.~Muraki$^{1}$, P.~Papini$^{2}$, A-L.~Perrot$^{9}$, S.~Ricciarini$^{2,4}$, T.~Sako$^{1,6}$, N.~Sakurai$^{6}$, Y.~Shimizu$^{7}$, T.~Suzuki$^{7}$, T.~Tamura$^{10}$, S.~Torii$^{7}$, A.~Tricomi$^{11,12}$, and W.C.~Turner$^{13}$\\
\llap{$^{1}$}Solar-Terrestrial Environment Laboratory, Nagoya University, Japan\\
\llap{$^{2}$}INFN Section of Florence, Italy \\
\llap{$^{3}$}University of Florence, Italy\\
\llap{$^{4}$}IFAC-CNR, Italy\\
\llap{$^{5}$}Ecole-Polytechnique, France\\
\llap{$^{6}$}Kobayashi-Maskawa Institute for the Origin of Particles and the Universe, Nagoya University, Japan\\
\llap{$^{7}$}RISE, Waseda University, Japan\\
\llap{$^{8}$}Graduate school of Science, Nagoya University, Japan \\
\llap{$^{9}$}CERN, Switzerland\\
\llap{$^{10}$}Kanagawa University, Japan\\
\llap{$^{11}$}INFN Section of Catania, Italy\\
\llap{$^{12}$}University of Catania, Italy\\
\llap{$^{13}$}LBNL, Berkeley, USA\\
E-mail:\email{kawade@stelab.nagoya-u.ac.jp}
}

\abstract{
The Large Hadron Collider forward (LHCf) experiment has been designed to use the LHC to benchmark the hadronic interaction models used in cosmic-ray physics.
The LHCf experiment measures neutral particles emitted in the very forward region of LHC collisions.
In this paper, the performances of the LHCf detectors for hadronic showers was studied with MC simulations and beam tests.
The detection efficiency for neutrons is from 60\% to 70\% above 500 GeV.
The energy resolutions are about 40\% and the position resolution is 0.1 to 1.3mm depend on the incident energy for neutrons.
The energy scale determined by the MC simulations and the validity of the MC simulations were examined using 350 GeV proton beams at the CERN-SPS.
}

\begin{document}
\maketitle
\flushbottom

\section{The LHCf experiment}
The LHCf experiment is dedicated to benchmark hadronic interaction models used in cosmic-ray physics \cite{bib:LHCfJINST,bib:LHCfTDR}.
The LHCf experiment is designed to measure particles emitted at the very forward region (pseudo-rapidity $\eta>8.4$) of the LHC.
We have already published photon and $\pi ^{0}$ spectra at 7TeV proton-proton collisions \cite{bib:LHCfphoton,bib:LHCfpizero} and photon spectra at 0.9TeV proton-proton collisions\cite{bib:LHCf900GeV}.
On the other hand, forward neutron measurement is also important to verify the hadronic interaction models.
However, the performances of the LHCf detectors for hadronic showers were not studied in detail.

Two independent detectors named ``Arm1'' and ``Arm2'' were installed in the instrumentation slots of the TANs (Neutral Target Absorbers) located $\pm$ 140m from IP1 of the LHC.
The Arm1 detector was installed on the IP8 side, and the Arm2 detector was installed on the IP2 side.
Only neutral particles (photons, $\pi^0$s, and neutrons) can be measured since charged particles are swept away by the D1 bending magnets between the interaction region IP1 and the front of the TANs.
Both Arm1 and Arm2 have two calorimeter towers.
Each calorimeter is composed of 16 layers of sampling scintillation panels interleaved with tungsten plates and position sensitive detectors.
The total depth of the calorimeters is 44 radiation lengths (1.6 hadron interaction lengths).
The transverse dimensions of the calorimeters are 20 mm $\times$ 20 mm and 40 mm $\times$ 40 mm in Arm1, and 25 mm $\times$ 25 mm and 32 mm $\times$ 32 mm in Arm2.
The smaller calorimeters in both arms are called ``Small towers'' and the larger called ``Large towers''.
The calorimeters were installed so that zero degree was covered by the Small towers.
Four pairs of X-Y position sensors were inserted between the sampling layers.
Scintillation Fiber (SciFi, 1 mm $\times$ 1 mm) detectors and silicon strip sensors (160 $\mu$m read-out pitch) are used for the position sensors of Arm1 and Arm2, respectively.
The details of the LHCf detectors have been previously reported \cite{bib:LHCfJINST,bib:LHCfTDR,bib:LHCfIJMPA}.

The detector performance for photon measurements has been reported previously \cite{bib:Mase}.
In this paper, the performances of the LHCf detectors for neutron measurement was studied with MC simulations in Section \ref{sec:MC} and beam tests in Section \ref{sec:SPS}.

\section{Performance study based on MC simulation up to 3.5 TeV hadron incident}\label{sec:MC}
\subsection{Monte Carlo setup}
The performances of the LHCf detectors for neutrons has been studied with Monte-Carlo simulations using the COSMOS (v7.49) and EPICS (v8.81) \cite{bib:EPICS} libraries that are used in air shower and detector simulations.
In addition, QGSJET II-03 \cite{bib:QGSJET2} has been used as a hadronic interaction model in the detector simulations for particle energies greater than 90 GeV and DPMJET3 \cite{bib:DPMJET3} for particle energies less than 90 GeV.
In order to estimate the dependence of the detector simulation on the choice of the model, we compared the results obtained by the QGSJET II + DPMJET3 (called QGSJET II, hereafter) model and the DPMJET3 model only as discussed in Section \ref{sec:SPS}.
In order to estimate energy and position resolutions and detection efficiency of the LHCf detectors, neutrons with energy in the range 100 GeV to 3500 GeV were injected to the center of the small tower and the large tower.
Neutrons with energy of 1 TeV were uniformly injected to the calorimeter to estimate the dependence of the shower leakage effects.

\subsection{Results}%
\subsubsection*{Detection efficiency}
\begin{figure}[]
  \centering
  \includegraphics[width=0.5\textwidth]{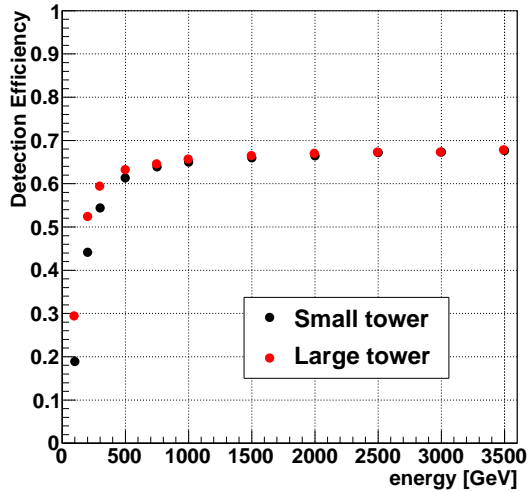}
  \caption{Detection efficiencies for neutrons as a function of energy for Arm1. The black (red) symbols correspond to the efficiencies of the Arm1 small (large) tower.}
  \label{detEFF}
 \end{figure}

Since the depth of the LHCf detectors is only 1.6 hadron interaction lengths, some neutrons pass through the detectors without interaction.
An offline event selection criterion was applied for neutron analysis in addition to an experimental trigger condition.
The neutron events were selected for analysis when the energy deposited in three successive scintillation layers exceeds the energy deposited by 200 MIPs (1 MIP is defined as 0.453 MeV).
The detection efficiencies for neutrons are shown in Figure \ref{detEFF} as a function of neutron energy for the Arm1 detector.
The black and red symbols correspond to the efficiencies for the Arm1 small and large towers, respectively. 
After offline event selection, the detection efficiency plateaus at a nearly constant 60\% to 70\% above 500 GeV.

\subsubsection*{Energy response and linearity}
The total energy deposited in the calorimeter was used as an estimator of the incident neutron energy.
An energy estimator named sumdE is defined as,
\begin{eqnarray*}
 {\rm sumdE} = \sum_{i=2}^{15} n_{step} \times dE_{i} 
\end{eqnarray*}
where $dE_{i}$ are energy deposited in the i-th sampling layer and $n_{step}$ are chosen as 1 for the 2nd to 10'th layers, and 2 for the 11'th to the last layers (proportional to the thickness of the tungsten).

\begin{figure}[]
 \centering
 \includegraphics[width=0.45\textwidth]{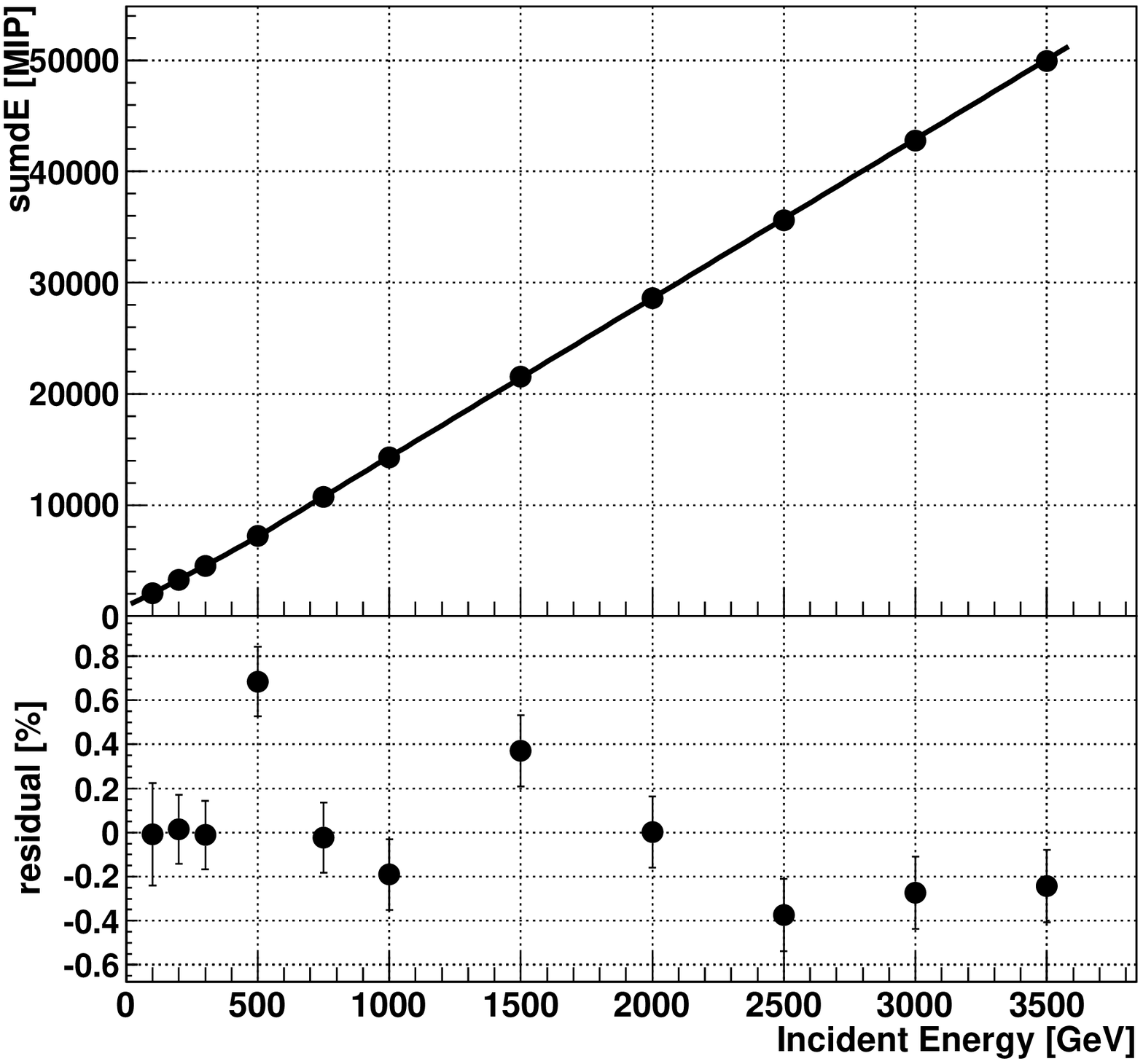}
 \includegraphics[width=0.45\textwidth]{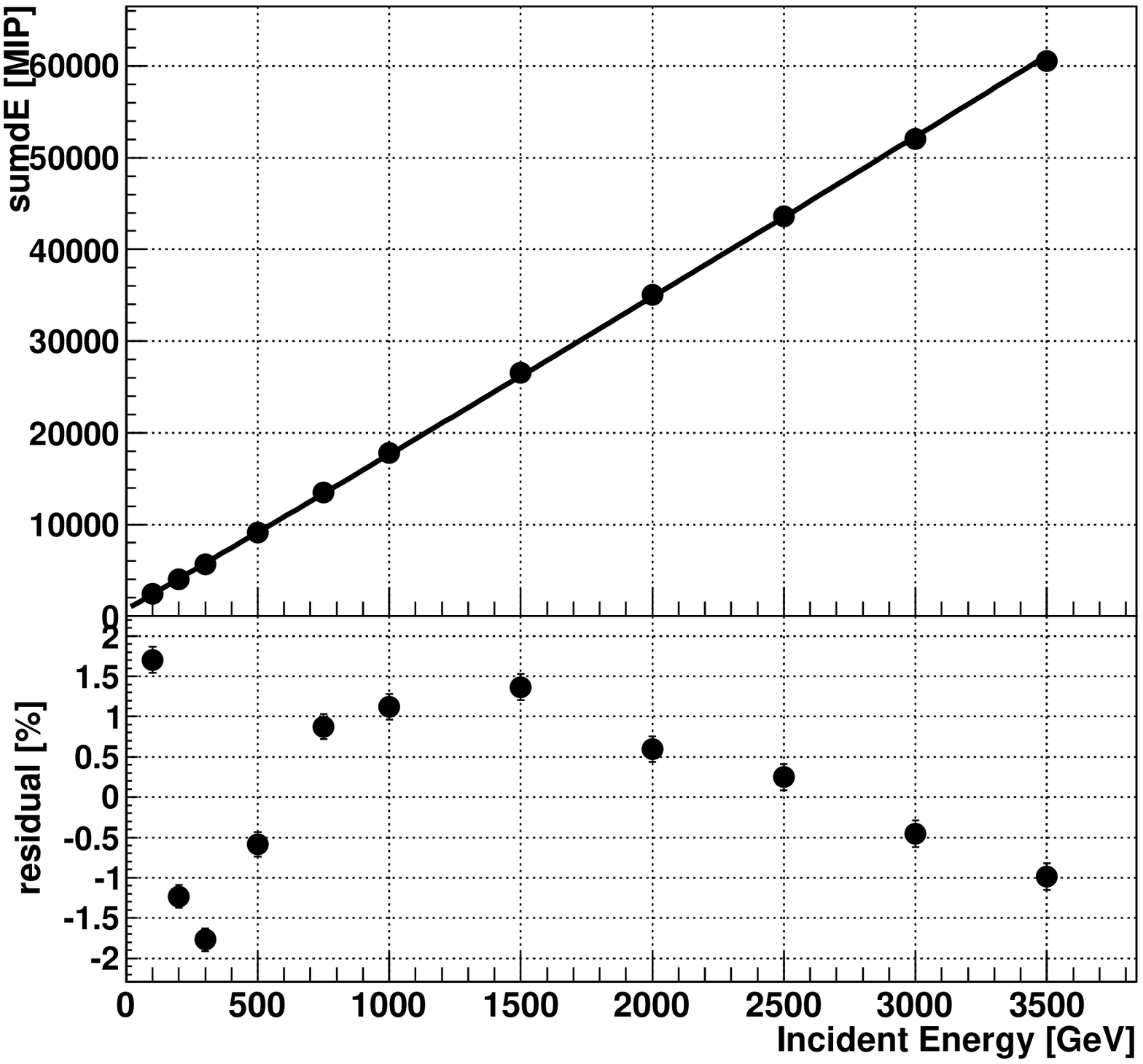}
 \caption{The energy response function for the Arm1 detector obtained by MC simulation with QGSJET II. The left figure corresponds to the small tower and the right corresponds to the large tower. The upper panels show the relations between the incident energy and the average of sumdE. The black curves are the results of fitting. The bottom panels show the residuals from the fitting with the functions shown in Eq. 2.1 and 2.2.}
 \label{fig:Linearity}
\end{figure}

\begin{figure}[]
 \centering
 \includegraphics[width=0.45\textwidth]{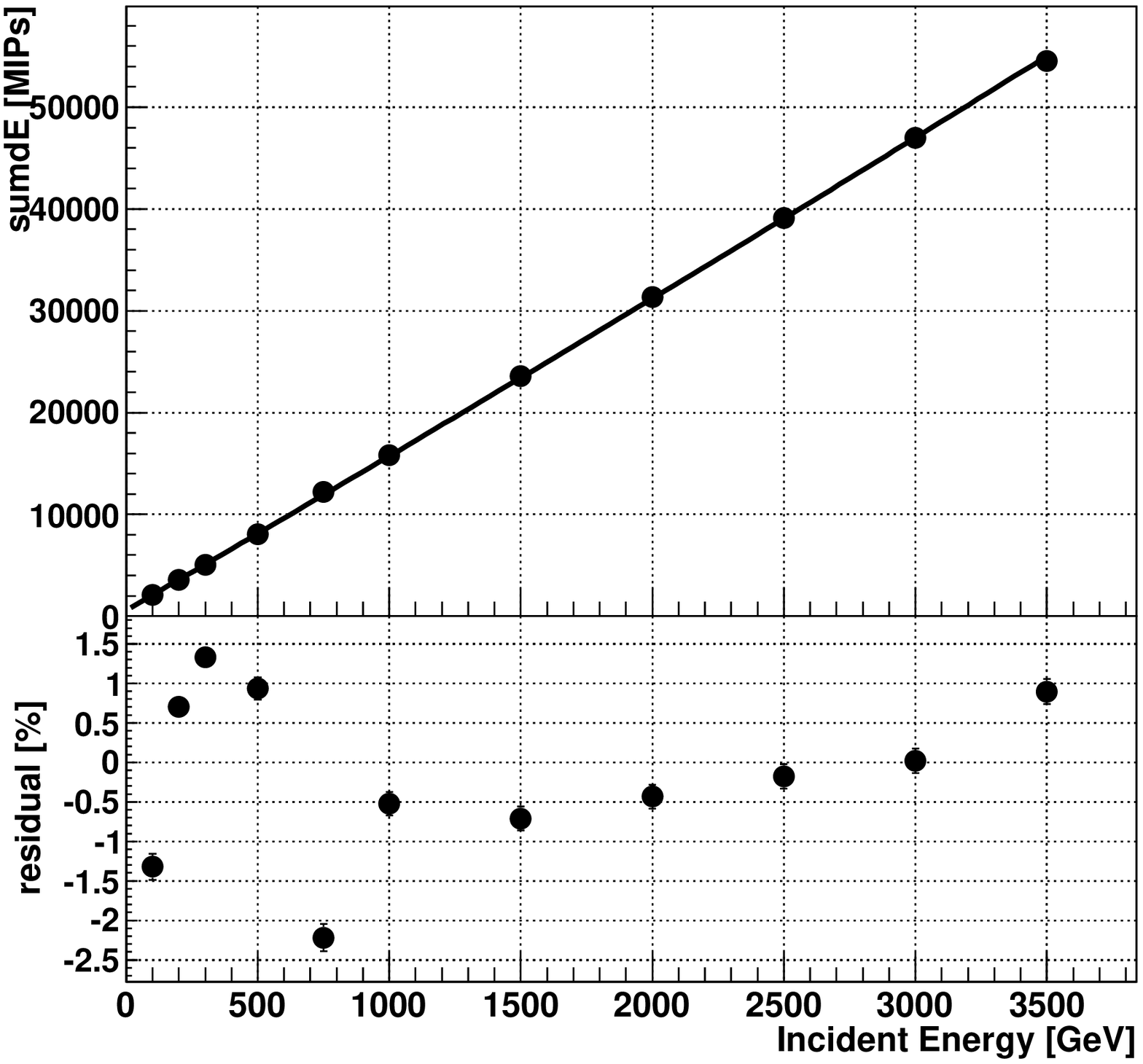}
 \includegraphics[width=0.45\textwidth]{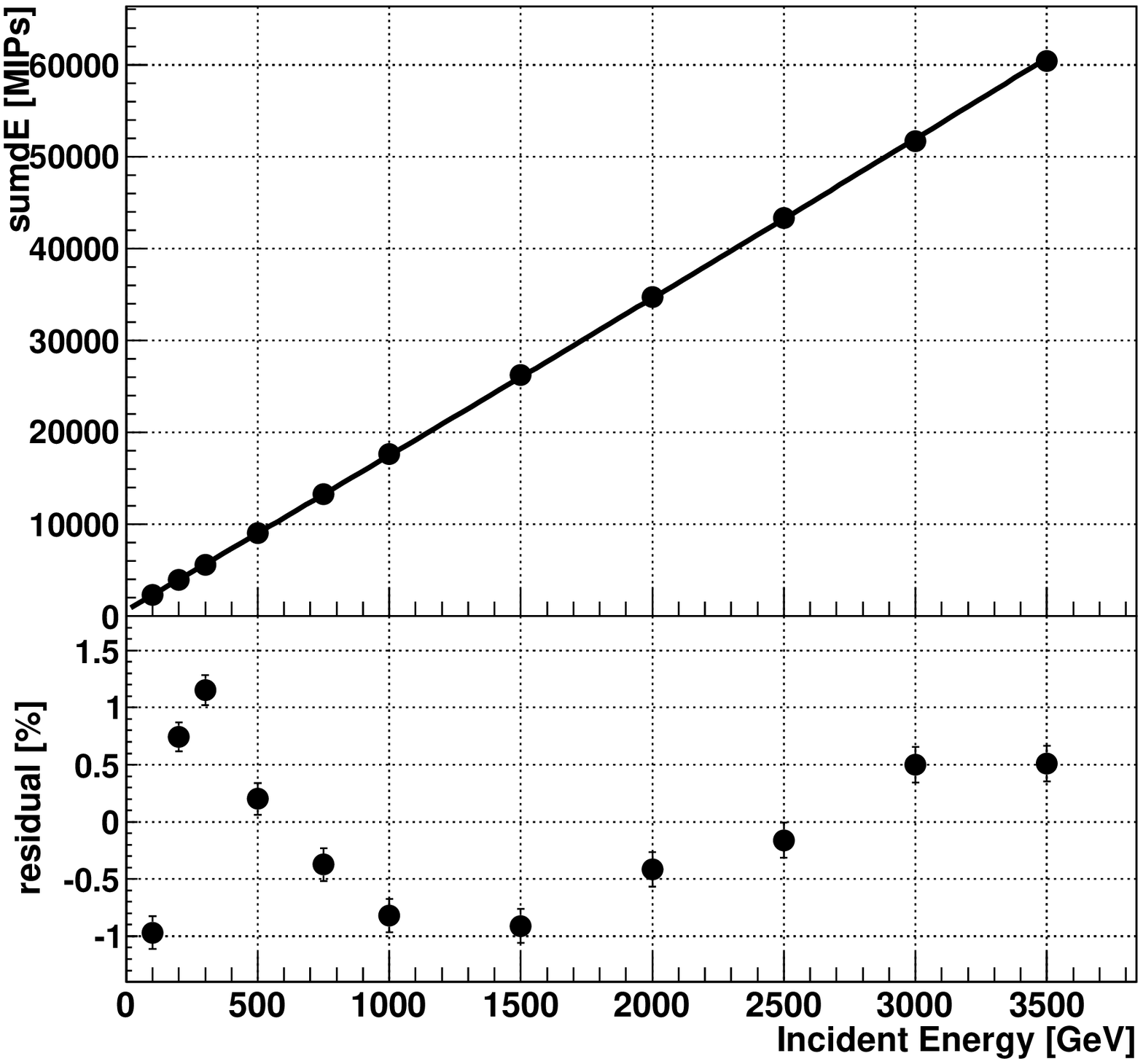}
 \caption{The energy response function for the Arm2 detector obtained by MC simulation with QGSJET II. The left figure corresponds to the small tower and the right corresponds to the large tower. The upper panels show the relations between the incident energy and the average of sumdE. The black curves are the results of fitting. The bottom panels show the residuals from the fitting with the functions shown in Eq. 2.1.}
 \label{fig:LinearityArm2}
\end{figure}

An energy response function was determined by the relationship between the incident energy E and sumdE for each tower as shown in Figures \ref{fig:Linearity} and \ref{fig:LinearityArm2}.
Neutrons with 100 GeV to 3500 GeV were injected at the center of the calorimeters for this study.
The horizontal axis corresponds to the energy of incident neutrons ($E$) and the vertical axis represents the average value of sumdE.
The response functions were derived by fitting the MC data with empirical polynomials.
A quadratic function (\ref{eq:qua}) was used as the energy response function for all the towers except the Arm1 small tower;
\begin{equation}\label{eq:energy}
 {\rm sumdE} = f(E) = \alpha E^2 + \beta E + \gamma .
 \label{eq:qua}
\end{equation}
For the Arm1 small tower, the function (\ref{eq:lin}) given below was used;
\begin{equation}\label{eq:energy2}
 {\rm sumdE} = f(E) = \left\{ \begin{array}{ll}
    \alpha E^2 + \beta E + \gamma & ({\rm E < 500GeV}) \\
    \delta E + \epsilon & ({\rm 500GeV < E }).  
  \end{array} \right .
  \label{eq:lin}
\end{equation}
The parameters expressed by the function (\ref{eq:lin}) were chosen to smoothly connect the data at 500 GeV.
The bottom panel of Figures \ref{fig:Linearity} and \ref{fig:LinearityArm2} represent the residuals from the fitting.
The non-linearity was less than $\pm$2\% for all the towers.
The error bars indicate the statistical uncertainty only.

\subsubsection*{Energy resolution}
\begin{figure}[]
 \centering
 \includegraphics[width=0.4\textwidth]{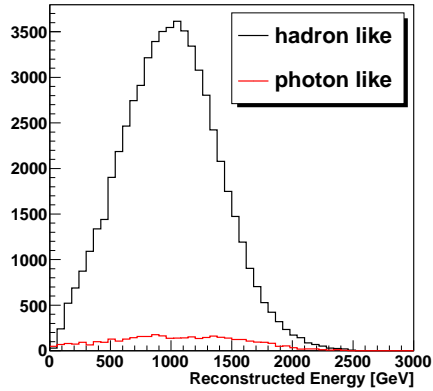}
 \caption{Results of MC calculation of reconstructed energy for 1 TeV incident neutrons for the Arm1 small tower. Red histogram shows neutrons misidentified as photons.}
 \label{Erec1TeV}
\end{figure}

The energy deposited by neutron induced showers has large fluctuations due to the limited length of the calorimeters.
Figure \ref{Erec1TeV} shows the distribution of the reconstructed energy for 1 TeV neutrons injected at the center of the small tower of Arm1.
The mean reconstructed energy of the distribution in Figure \ref{Erec1TeV} is 997 $\pm$ 2 GeV.

\begin{figure}[]
  \centering
  \includegraphics[width=0.4\textwidth]{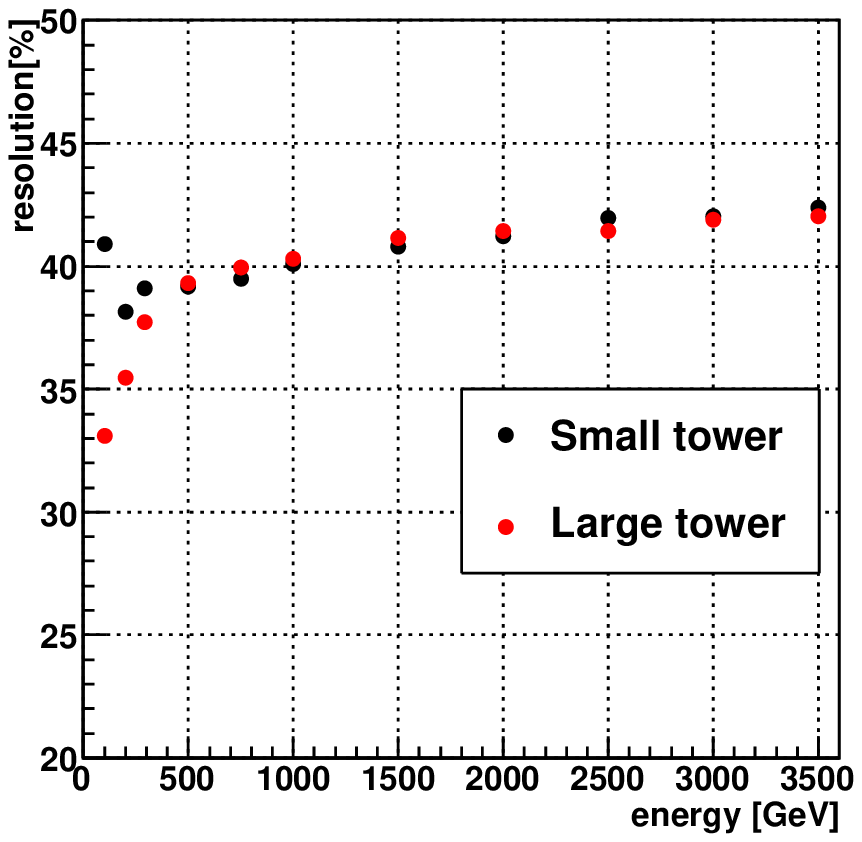}
  \includegraphics[width=0.4\textwidth]{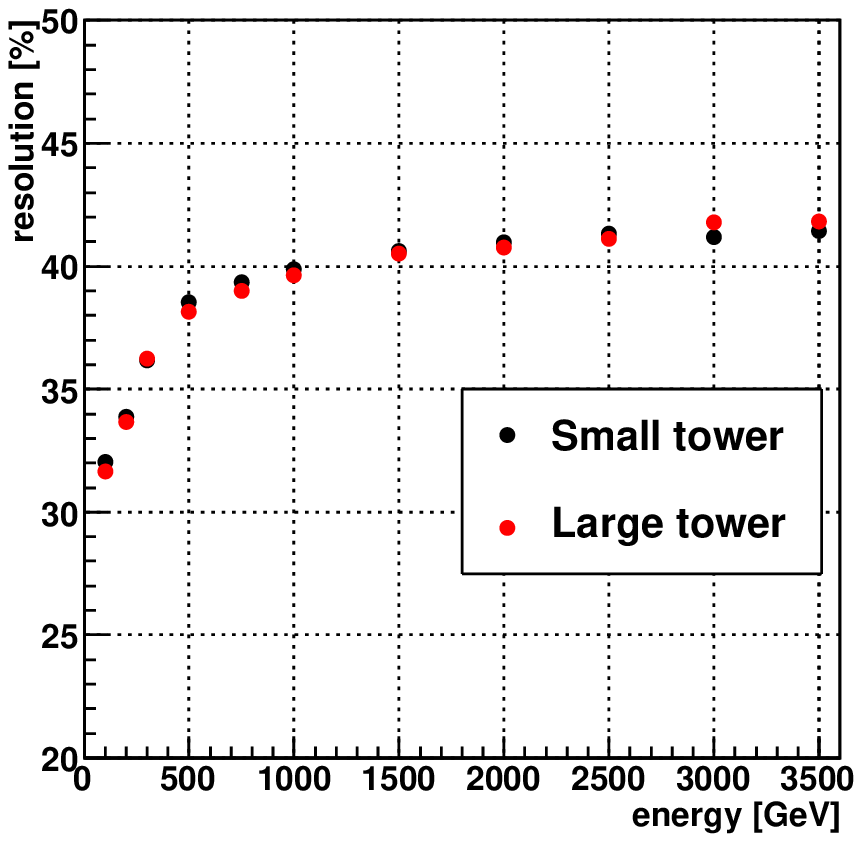}
  \caption{The energy resolution of the small tower (black) and the large tower (red) for neutrons as a function of energy. The left panel is for the Arm1 detector and the right panel for Arm2.}
  \label{EneRes}
 \end{figure}

Neutron-like or photon-like events were discriminated by using a PID (Particle Identification) algorithm as discussed in Section \ref{sec:PID}.
Because the visible energy of neutrons in the calorimeter is roughly 30\% of that for photons with same energy, the energy estimation depends on the PID result.
Although all the events in this MC study are neutron events, we selected only those that would pass the neutron-like selection in an experimental situation for calculation of the neutron energy resolution.
The black and red histograms in Figure \ref{Erec1TeV} correspond to neutron-like events and photon-like events, respectively.  
The energy resolution is defined as the standard deviation of the reconstructed energy distribution.
Figure \ref{EneRes} shows the expected energy resolutions of the small tower and the large tower as a function of the incident energy.
The left panel and right panel correspond to the energy resolutions of the Arm1 detector and the Arm2 detector, respectively.
From the results of these MC calculations, the energy resolutions are about 40\% for the all towers.

\subsubsection*{Transverse hit position resolution}
\begin{figure}[]
  \centering
  \includegraphics[width=0.4\textwidth]{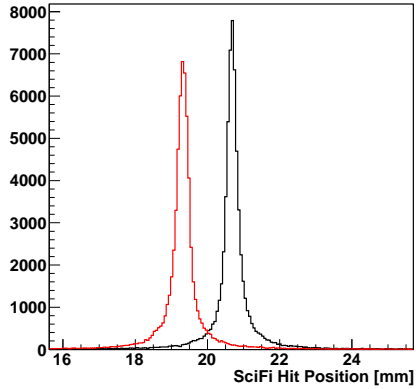}
  \caption{Reconstructed hit position distributions for the Arm1 large tower. The black and red curves correspond to the reconstructed transverse hit positions of X and Y axes, respectively.}
  \label{Postion1D}
 \end{figure}

\begin{figure}[]
  \centering
  \includegraphics[width=0.9\textwidth]{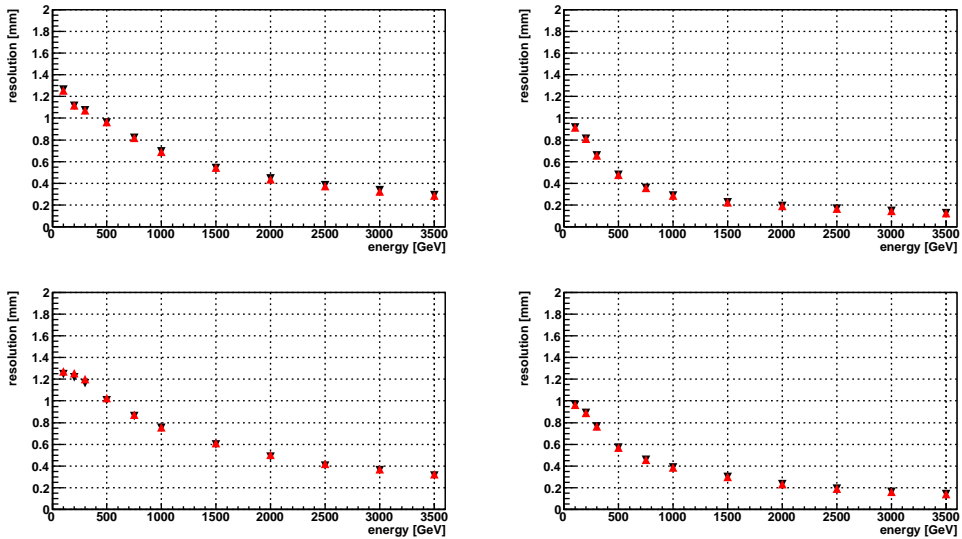}
  \caption{The position resolution for neutrons as a function of energy. Left (Right) panels correspond to the result of Arm1 (Arm2). Upper and lower panels correspond to the resolution of the small tower and the large tower, respectively. The black (red) markers indicate the transverse hit position resolutions of the X (Y) axis.}
  \label{PosRes}
 \end{figure}

The transverse hit position is one of the important observables not only for the determination of the transverse momentum p$_{{\rm T}}$ but also for the estimation of the energy.
Because the transverse sizes of the LHCf detectors are limited, a fraction of the shower particles leaks out of the calorimeters.
The effect of the leakage as discussed later is position dependent.
Firstly, the transverse hit positions were measured by the fitting method developed in a previous study of the four SciFi and silicon layers \cite{bib:LHCfphoton}.
The transverse hit position reconstructed for the layer which has the largest calorimetric signal was then selected.
Figure \ref{Postion1D} shows the distribution of the reconstructed transverse position when 1 TeV neutrons were injected into the center of the large tower (position X=20.8mm and Y=19.3mm in this coordinates).
The position resolution is defined as the Full Width of Half Maximum (FWHM) of the distribution. 
Figure \ref{PosRes} shows the expected resolution of transverse incident positions measured by the position sensitive layers of Arm1 and Arm2 detectors.
The position resolution varies from 0.1 to 1.3mm depend on the incident energy.

\subsubsection*{Shower leakage effects}
\begin{figure}[]
  \centering
  \includegraphics[width=0.35\textwidth]{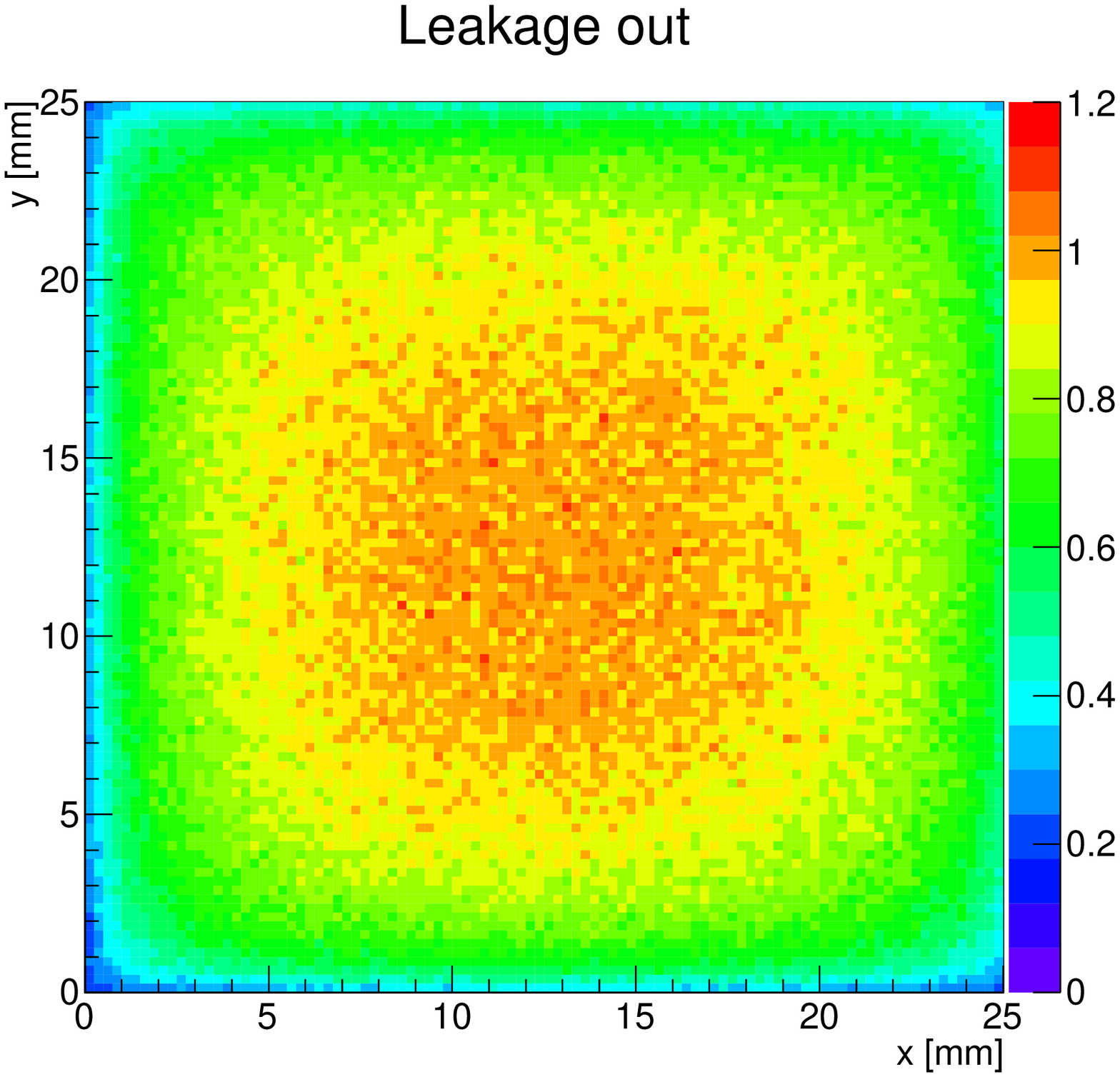}
  \includegraphics[width=0.35\textwidth]{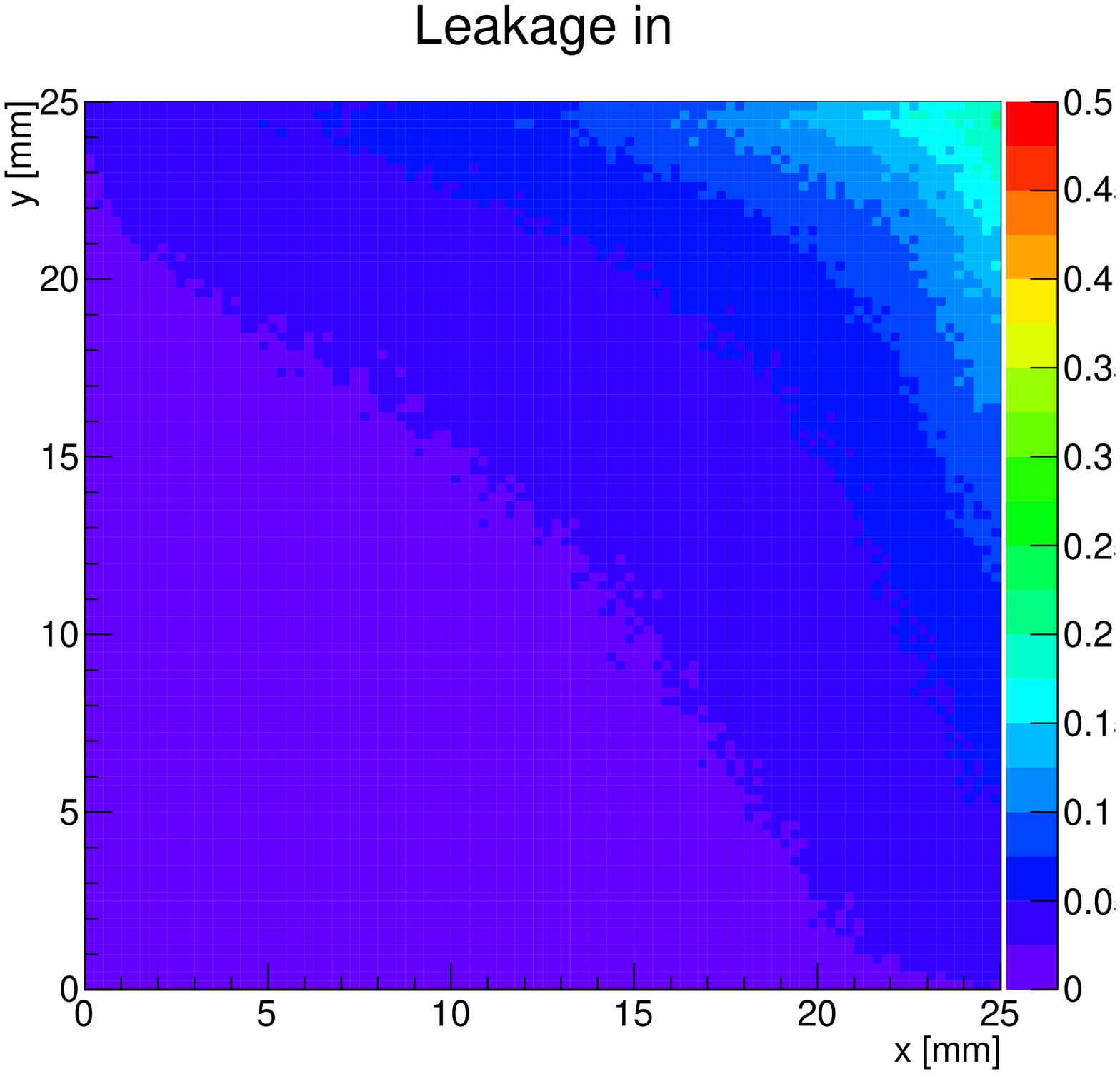}
  \caption{The left panel shows the fraction of energy deposited in the Arm2 small tower as a function of the X and Y coordinates of the incident neutrons. The color contours show deposited energy fractions relative to neutrons injected at the center of the calorimeter. The right panel shows the fraction of neutron energy deposited in the large tower of the Arm2 detector as a function of the X and Y coordinates of a neutron incident on the small tower of the Arm2 detector.}
  \label{Leakout}
 \end{figure}

As previously mentioned, a fraction of the shower particles leaks out from the LHCf calorimeters due to their limited size.
The effect of the leakage has been estimated by using MC simulations.
The left panel of Figure \ref{Leakout} shows the position dependence of mean sumdE, <sumdE>, in the Arm2 small tower.
The X and Y axes show the incident neutron position and color contours correspond to the value of sumdE relative to the value at the center of the detector. 
Neutrons with 1 TeV were injected uniformly and the deposited energy in each position were analyzed.
Because the energy resolution worsens significantly for the events hitting within 2mm from the edge of the calorimeters, they are removed from the analysis.

If one or more particles hits each calorimeter of the Arm1 or Arm2 detector at the same time, the shower leakages affect the energy measures in each calorimeter.
This effect is called shower leakage-in as some of the energy leaking out from one calorimeter leaks in to the neighbor one.
While leakage-in has a significant effect on the measurements of $\pi^{0}$ mesons \cite{bib:LHCfpizero}, it also has some effect on the neutron measurements.
The right panel of Figure \ref{Leakout} shows the fraction of neutron energy deposited in the large calorimeter of the Arm2 detector as a function of the X and Y coordinates of a neutron incident on the small tower of the Arm2 detector.
The leakage-in and the leakage-out effects were taken into account in the equations below.
\begin{equation}\label{eq:leak}
 M^{TS} = L_{out}^{TS}T^{TS} + L_{in}^{TL}T^{TL},
\end{equation} 
\begin{equation}\label{eq:leak2}
 M^{TL} = L_{in}^{TS}T^{TS} + L_{out}^{TL}T^{TL}
\end{equation} 

Here, $T^{TS}$ and $T^{TL}$ are the deposited energies (sumdE) when the particles hit the center of the small and large towers, respectively.~
$L_{out}$ and $L_{in}$ correspond to the leakage-out factor and leakage-in factor as functions of the hit position, respectively.~
$M^{TS}$ ($M^{TL}$) means the measured deposited energy in the small tower (large tower).

Because the energy resolution gets worse due to the leakage effect, we can correct this effect by applying leakage correction as functions of their hit positions using equation below.
\begin{eqnarray}
 T^{TS} = \frac{L_{out}^{TL}M^{TS} - L_{in}^{TL}M^{TL}}{L_{out}^{TS}L_{out}^{TL}-L_{in}^{TS}L_{in}^{TL}},\\
 T^{TL} = \frac{L_{out}^{TS}M^{TL} - L_{in}^{TS}M^{TS}}{L_{out}^{TS}L_{out}^{TL}-L_{in}^{TS}L_{in}^{TL}} 
\end{eqnarray}
They are the solution of the simultaneous equations \ref{eq:leak} and \ref{eq:leak2} and used as sumdE in the functions \ref{eq:energy} and \ref{eq:energy2}.

\section{The SPS beam test}\label{sec:SPS}
\subsection{The overview of the experiment}
The performance of the calorimeters for the measurement of hadronic showers was studied in 2007 and 2010 by using 350 GeV proton beams at the CERN-SPS H4 beam line.
The consistency of the MC simulations was also checked by comparing them with the results of the beam tests.
For the MC simulations, we used the COSMOS and EPICS packages as explained in Section \ref{sec:MC}.
Conversion of charge measured by the ADCs to the number of minimum ionizing shower particles (MIPs) was based on the conversion factors previously obtained by using electron and muon beams \cite{bib:Mase}.

\begin{figure}[]
 \centering
 \includegraphics[width=0.8\textwidth]{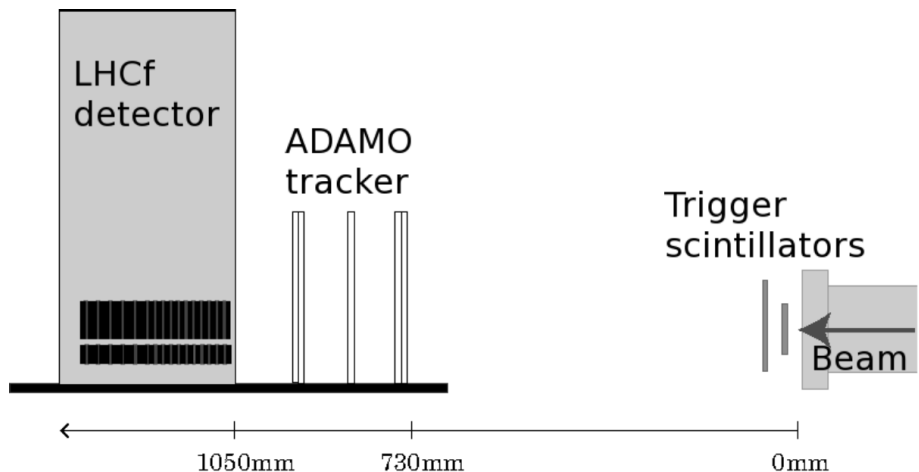}
 \caption{The setup of the SPS experiment. The trigger signals were generated by using small scintillators at the exit of beam pipe. The detector was mounted on a movable table together with the ADAMO tracker to scan the calorimeter through the beams.}
 \label{SPSsetup}
\end{figure}
Figure \ref{SPSsetup} shows the set-up of the SPS test beam experiment.
The trigger signals were generated by using small scintillators (20 mm $\times$ 20 mm and 40 mm $\times$ 40 mm) placed behind the thin beam exit window as shown in the figure.
Then, precise transverse hit positions of the test beam particles were measured by using the ADAMO tracker \cite{bib:ADAMO} installed in front of the LHCf detector.
The ADAMO tracker is composed of silicon strip sensors with position resolution less than 20 $\mu$m.
The ADAMO tracker has enough resolution to determine the position resolution of the LHCf detectors.
The LHCf detector and ADAMO tracker were mounted on a movable table to allow scanning the calorimeters through the test beams.
The beam position determined by the ADAMO tracker was used for analysis of the data.

\begin{figure}[]
 \centering
 \includegraphics[width=0.45\textwidth]{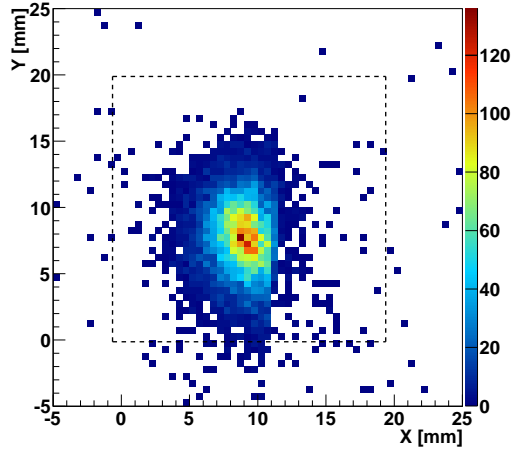}
 \caption{The beam profile of the 350 GeV proton test beam measured by the ADAMO tracker. Dashed line shows the edge of Arm1 small tower. Color contour shows the number of particles.}
 \label{SPSflux}
\end{figure}

Figure \ref{SPSflux} shows the transverse profile of the 350 GeV proton beam measured by the ADAMO tracker.
The X and Y axes represent the transverse position coordinates of the ADAMO tracker.
To simplify the analysis only the calorimeter leakage-out corrections were applied and the leakage-in corrections were ignored.
For this operation, the high voltages of the PMTs used in the LHCf calorimeters were set at the low gain mode (400-475V).
This is similar to the setting of the PMT high voltages during the operation at the LHC.

\subsection{Analysis and results}\label{sec:PID}
In this section the analysis results of the Arm1 detector are provided.
The results of Arm2 analysis are consistent with those of Arm1.
The data analysis was carried out in the following way.
The raw data (measured ADC counts) were converted into the numbers of minimum ionizing particles passing through the scintillation layers of the calorimeters after subtracting the pedestal.
The offline event selection which was more than 200MIPs for successively three continuous layers was applied.
Then the transverse hit position was reconstructed from the SciFi detector.
The parameters, L$_{20\%}$ and L$_{90\%}$ that represent the longitudinal development of showers \cite{bib:LHCfphoton}, were obtained from the shower transition shape.
Here, L$_{20\%}$ and L$_{90\%}$ parameters are the calorimeter depths containing 20\% and 90\% of the total deposited energy, respectively.
A two dimensional selection for L$_{20\%}$ and L$_{90\%}$ parameters was employed to perform PID more efficiently with less contamination. 
An optimized parameter L$_{{\rm 2D}}$ defined as ${\rm L}_{\rm 2D} = {\rm L}_{90\%} - 1/4 \times {\rm L}_{20\%} $ was introduced.
Hadronic showers are more penetrating than electromagnetically induced showers so contamination by electrons would show up as an excess of events with low values of the L$_{{\rm 2D}}$ parameter.

\begin{figure}[]
 \centering
 \includegraphics[width=0.5\textwidth]{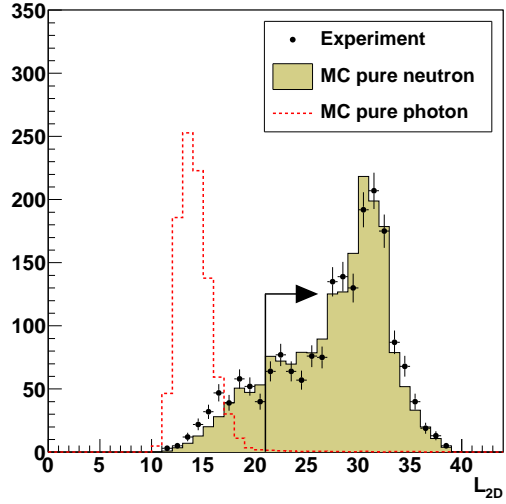}
 \caption{Comparison of the L$_{2D}$ distribution for the 350 GeV proton test beam and the MC prediction for the Arm1 small tower. Black crosses show the experimental data and the histogram shows the MC prediction. Dashed red line indicates L$_{2D}$ distribution of pure photons. Arrow shows the PID selection criteria in this analysis.}
 \label{SPS_L90}
\end{figure}
Figure \ref{SPS_L90} shows the distribution of the L$_{{\rm 2D}}$ parameter for the experimental data (black marker) and the MC simulation (Filled histogram).
Dashed red line shows L$_{{\rm 2D}}$ distribution of pure photons.
The arrow at L$_{{\rm 2D}}$ = 21 indicates PID selection criteria.
The agreement between the MC prediction and the experimental data is quite satisfactory ($\chi^{2}$/NDF is about 31.5/28 taking into account the statistical uncertainties), so there seems to be no significant contamination in the proton beam by positrons.

\begin{figure}[]
 \centering
 \includegraphics[width=0.8\textwidth]{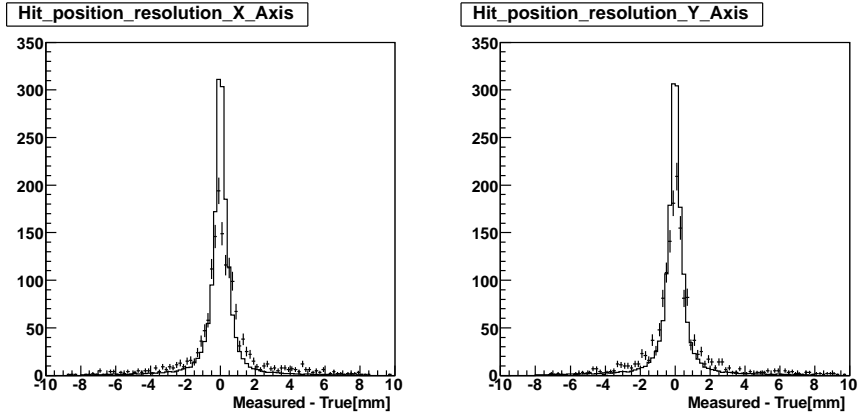}
 \caption{The difference between the SciFi measured hit position and ``true'' hit position measured by the ADAMO tracker in the case of a 350 GeV proton beam for the Arm1 small tower. The left(right) panel is the result for the X (Y) axis. The data points correspond to the experimental data, while the histograms correspond to the MC simulations. For the experimental data, the hit position determined by the ADAMO tracker used as true hit position.}
 \label{fig:posres}
\end{figure}

The performance of position determination was estimated by comparing the hit position determined by the SciFi detector with the ADAMO tracker.
Figure \ref{fig:posres} shows the difference between the SciFi measured hit position and the beam hit position determined by the ADAMO tracker.
The left (right) panel shows the results for the X (Y) coordinate.
The data points correspond to the experimental data while the histograms show the MC simulations.
In the neighborhood of the peaks the data and MC simulations are consistent but the experimental data contain a tail component which does not appear in the simulations.

The energy deposited in each scintillation layer was compared with the MC prediction.
To reproduce the actual experimental conditions, the measured pedestal fluctuations were included in the MC simulations.
\begin{figure}[t]
 \centering
 \includegraphics[width=1.0\textwidth]{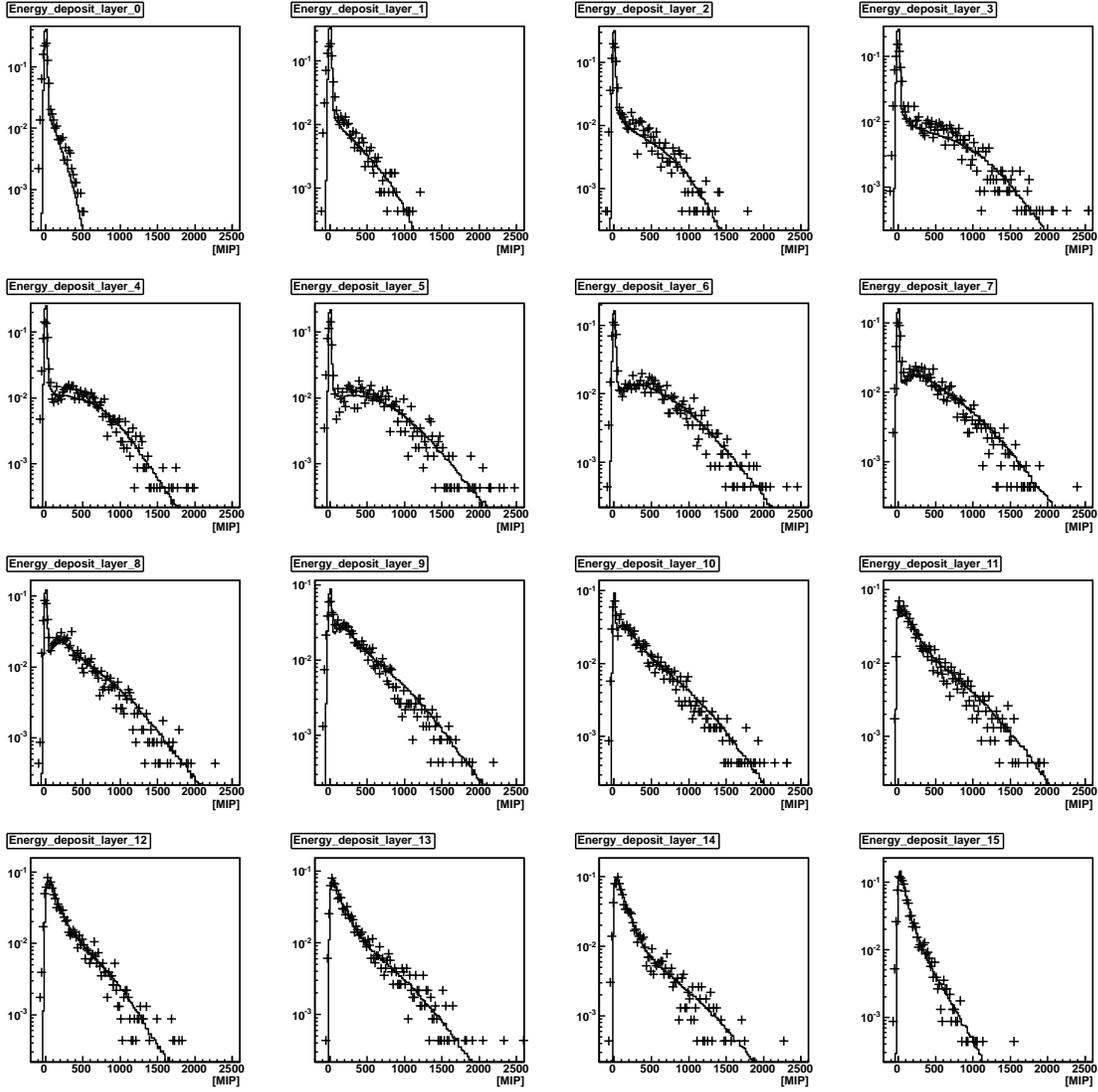}
 \caption{The deposited energy distribution in each scintillation layer of the Arm1 small tower for the 350 GeV SPS proton test beam. The black data points indicate the experimental results. For comparison MC simulations for 350 GeV neutrons are overlaid as solid black line histogram.}
 \label{SPS350GeV}
\end{figure}
Figure \ref{SPS350GeV} shows the deposited energy distribution in the scintillation layers of the small tower for the 350 GeV proton beam data and for 350 GeV neutrons simulated by MC calculations.
The difference between the data and the MC prediction is summarized in Table \ref{tab:ffac}.
\begin{table}[!h]
  \begin{center}
   \caption{The ratios of measured energy deposited by 350 GeV SPS protons to the MC simulation of energy deposited by 350 GeV neutrons for all the layers of Arm1 small tower.}
    \begin{tabular}{cc|cc|cc|cc}
     Layer & Ratio & Layer & Ratio & Layer & Ratio & Layer & Ratio\\ \hline
     0 & 1.1577 & 4& 1.0370 & 8  & 0.8726 & 12 & 0.9064 \\ 
     1 & 0.9998 & 5& 0.9965 & 9  & 0.8483 & 13 & 0.9785 \\ 
     2 & 1.0336 & 6& 0.9625 & 10 & 0.9117 & 14 & 0.8842 \\ 
     3 & 1.0262 & 7& 0.8904 & 11 & 0.8582 & 15 & 0.8942 \\ 
    \end{tabular}
   \label{tab:ffac}
  \end{center}
\end{table}
The Ratio in i-th layer $r^{i}$ was calculated by,
\begin{equation}
 r^{i} = {\rm <}dE_{Ex}^{i}{\rm >} / {\rm <}dE_{MC}^{i}{\rm >},
\end{equation}
where <$dE_{EX}^{i}$> and <$dE_{MC}^{i}$> are the average energy deposit in the i-th layer for the experiment and the MC, respectively.

Possible reason for the differences in the $r^{i}$ from unity are uncertainty of the hadronic interaction length of the tungsten plates and the gain calibrations performed in the previous study \cite{bib:Mase}.
\begin{figure}[]
 \centering
 \includegraphics[width=0.45\textwidth]{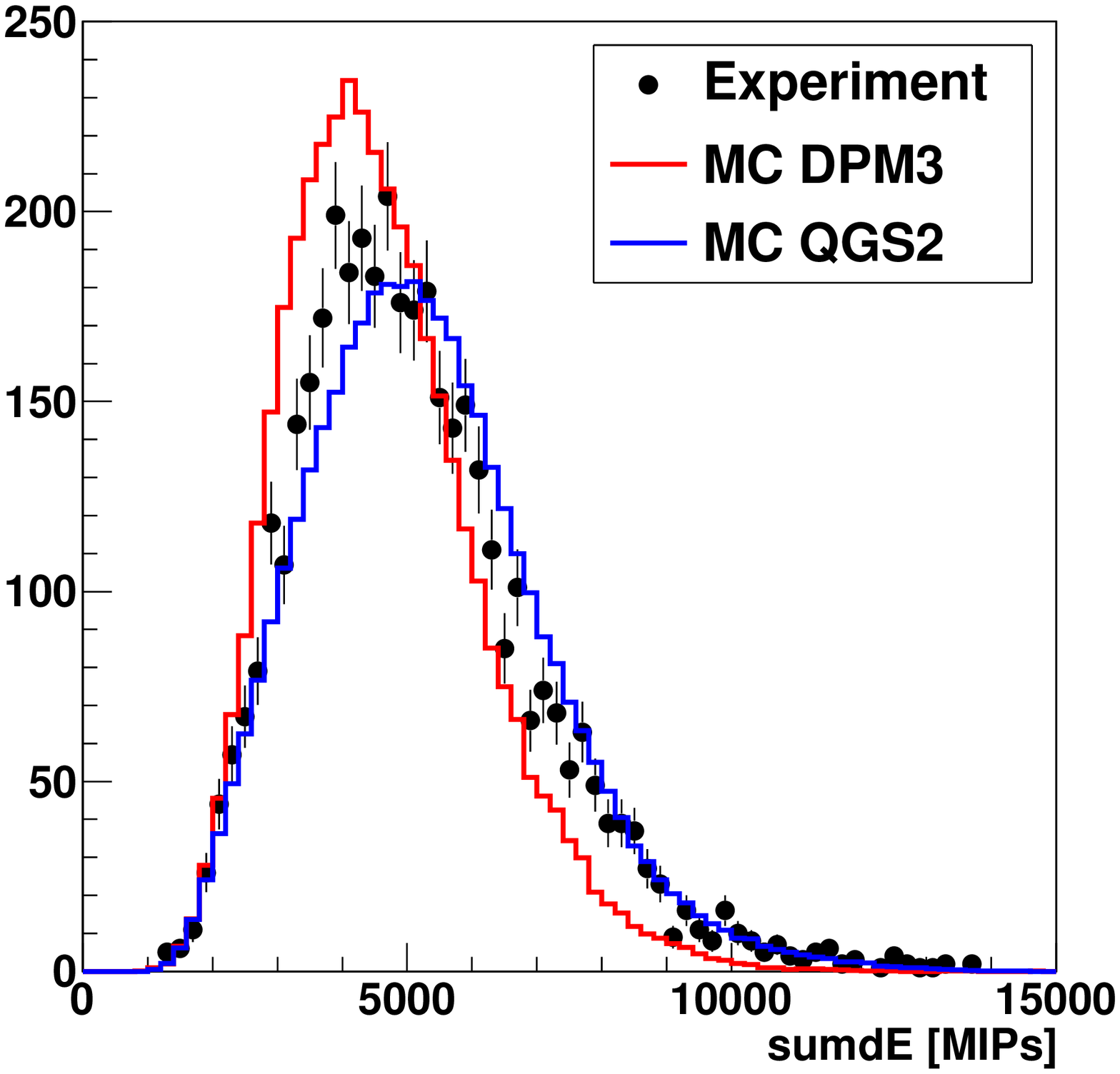}
 \includegraphics[width=0.45\textwidth]{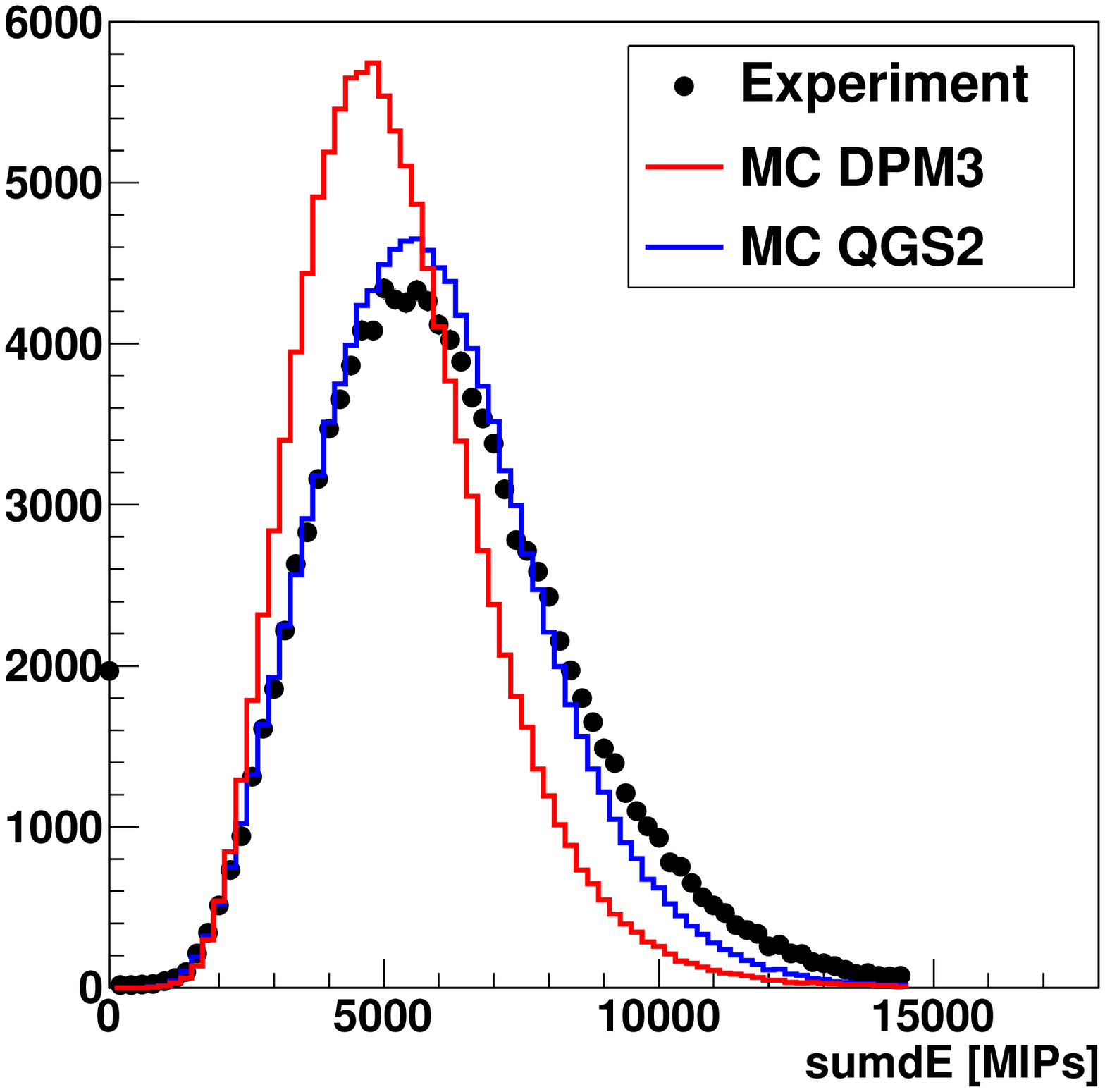}
 \caption{Comparison of the sumdE distributions for the experimental 350 GeV proton data and 350 GeV neutron MC simulations. The black symbols are the experimental data. The red and the blue histograms represent the MC results with the DPMJET3 model and the QGSJET II model, respectively. The left (right) figure corresponds to the result of the Arm1 (Arm2) small tower.}
 \label{SPSsumdEcomp}
\end{figure}
Figure \ref{SPSsumdEcomp} presents comparison of the sumdE distribution for the experimental 350 GeV SPS proton beam data and MC simulation of 350 GeV neutrons.
The black symbols are the experimental data while the red and blue histograms are the MC predictions with the DPMJET3 model and the QGSJET II model, respectively.
The mean, standard deviation ($\sigma$) of sumdE and their ratio ($\sigma$/mean) are summarized in Table \ref{tab:SPS}.
The ratios of the mean values in MC with respect to the experimental data are also shown.
\begin{table}[!h]
  \begin{center}
   \caption{Mean, $\sigma$, $\sigma$/Mean and Ratio (MC/Experiment) of sumdE distributions for the experimental 350 GeV proton data and the MC results for 350 GeV neutrons with the DPMJET3 and QGSJET II hadronic interaction models.}
    \begin{tabular}{cl|cccc}
        &              & Mean [MIPs] & $\sigma$  & $\sigma$/Mean & Ratio \\ \hline
        & Experiment   & 5116        & 1840 & 0.360 & -    \\ 
  Arm1  & MC DPMJET3   & 4590        & 1484 & 0.323 & 0.897    \\ 
        & MC QGSJET II & 5294        & 1822 & 0.341 & 1.035    \\ \hline
        & Experiment   & 6022        & 2379 & 0.395 & -    \\ 
  Arm2  & MC DPMJET3   & 4968        & 2010 & 0.405 & 0.825    \\ 
        & MC QGSJET II & 5631        & 2284 & 0.406 & 0.935    \\ 
    \end{tabular}
   \label{tab:SPS}
  \end{center}
\end{table}
The QGSJET II model has mean and standard deviation of sumdE similar to the experimental data.
On the other hand, the DPMJET3 model clearly underestimates the width of the distribution, especially at the high energy tail.
The +3.5\% (-6.5\%) difference in sumdE for Arm1 (Arm2) between the experimental 350 GeV proton data and the MC simulation with the QGSJET II model will be included as part of the systematic uncertainties when we reconstruct the incident energy based on the functions 2.1 and 2.2.

\section{Conclusions}
The performances of the LHCf detectors for hadronic showers were studied by using 350 GeV proton test beams and MC simulations.
The detection efficiency is greater than 60\% for neutrons above 500 GeV for both towers of the Arm1 and Arm2 detectors.
The energy resolution is about 40\% above 500 GeV and weakly dependent on energy.
The $\pm$2\% non-linearity of the energy scale is confirmed.
The resolution of transverse hit position is less than 1 mm above 500 GeV and decreases slowly with increasing energy.
The performances of the detectors for hadronic showers and the validity of the MC simulations were tested using the 350 GeV proton beams at CERN SPS in 2007 and 2010.
We compared two different MC configurations using DPMJET3 or QGSJET II hadronic interaction models.
The QGSJET II model shows better agreement with the experimental test beam data.
The 36\% sumdE resolution of experimental data is consistent with the prediction of MC simulations as shown in Figure \ref{EneRes}. 
The energy scale was checked by comparing the sumdE distribution of the data with the MC simulations.
The 3.5\% and 6.5\% difference in sumdE between the experimental result and the MC prediction with the QGSJET II model will be included as part of the systematic uncertainty.
The experimental and MC results for the Arm1 and Arm2 detectors are consistent.

\section*{Acknowledgments}
We would like to express our appreciation to the CERN SPS staff for their contribution to the work reported in this paper.
This study was supported by Grant-in-Aids for Scientific Research by MEXT of Japan, by the Grant-in-Aid for Nagoya University GCOE ``QFPU'' from MEXT and by Istituto Nazionale di Fisica Nucleare (INFN) in Italy.
In addition part of this work was performed using computer resources provided by the Institute for Cosmic-Ray Research at the University of Tokyo and by CERN.

\end{document}